\documentclass[runningheads]{llncs}

\usepackage{times}
\usepackage{graphicx}
\usepackage[font=small,skip=0pt]{subcaption}  
\usepackage{ulem}  
\usepackage{amsmath}

\usepackage[utf8]{inputenc} 
\usepackage[T1]{fontenc}    
\usepackage{hyperref}       
\usepackage{url}            
\usepackage{booktabs}       
\usepackage{amsfonts}       
\usepackage{nicefrac}       
\usepackage{microtype}      
\usepackage[perpage]{footmisc}  

\usepackage[numbers,sectionbib,sort&compress]{natbib}

\title{Multi-Channel Stochastic Variational Inference for the Joint Analysis of Heterogeneous Biomedical Data in Alzheimer's Disease}
\titlerunning{Multi-Channel Variational Inference}

\author{
	Luigi Antelmi\inst{1} \and Nicholas Ayache\inst{1} \and Philippe Robert\inst{2,3} \and Marco Lorenzi\inst{1} \\
	for the  Alzheimer's  Disease  Neuroimaging  Initiative
 	\thanks{
 		Data used in preparation of this article were obtained from the Alzheimer's Disease Neuroimaging Initiative (ADNI) database (\href{http://adni.loni.usc.edu}{adni.loni.usc.edu}).
 		As such, the investigators within the ADNI contributed to the design and implementation of ADNI and/or provided data but did not participate in analysis or writing of this report.
 		A complete listing of ADNI investigators can be found at: \href{http://adni.loni.usc.edu/wp-content/uploads/how_to_apply/ADNI_Acknowledgement_List.pdf}{http://adni.loni.usc.edu/wp-content/uploads/how\_to\_apply/ADNI\_Acknowledgement\_List.pdf}.
 	}
}

\institute{
	University of Côte d'Azur, Inria Sophia Antipolis, Epione Research Project, France
	\and
	University of Côte d'Azur, CoBTeK, France
	\and
	Centre Memoire, CHU of Nice, France
}

\authorrunning{L. Antelmi et al.}

\newcommand{\figref}[1]{Fig.~\ref{#1}}
\newcommand{\tabref}[1]{Tab.~\ref{#1}}
\newcommand{\eqnref}[1]{Eq.~(\ref{#1})}

\newcommand{\cf}[0]{\textit{cf. }}
\newcommand{\eg}[0]{\textit{e.g., }}

\newcommand{\ie}[0]{\textit{i.e.,}}

\newcommand{\asuppmat}[0]{\hyperref[sec:supmat]{\textit{Sup. Mat.}} }

\newcommand{\rhs}[0]{right hand side}  
\newcommand{\lhs}[0]{left hand side}  
\newcommand{\snr}[0]{snr}  
\newcommand{\realnumbers}[0]{\mathbb{R}}

\newcommand{\xb}[0]{\mathbf{x}}
\newcommand{\setx}[0]{\mathbf{x}_1,\ldots,\mathbf{x}_C}

\newcommand{\zb}[0]{\mathbf{z}}
\newcommand{\dz}[0]{d\zb}

\newcommand{\thetab}[0]{\boldsymbol{\theta}}
\newcommand{\thetaset}[0]{\thetab_1,\ldots,\thetab_C}
\newcommand{\phib}[0]{\boldsymbol{\phi}}

\newcommand{\epsilonb}[0]{\boldsymbol{\epsilon}}
\newcommand{\x}[0]{\xb}
\newcommand{\z}[0]{\zb}
\newcommand{\q}[1]{q\left(#1\right)}
\newcommand{\p}[1]{p\left(#1\right)}

\newcommand{\Gauss}[2]{\mathcal{N}\left(#1;#2\right)}
\newcommand{\Gaussof}[1]{\mathcal{N}\left(#1\right)}
\newcommand{\Gausstd}[0]{\Gauss{\mathbf{0}}{\mathbf{I}}}
\newcommand{\GaussStdDim}[1]{\Gauss{\mathbf{0}}{\mathbf{I}_{#1}}}
\newcommand{\KL}[2]{\mathcal{D}_{KL}\big(#1\left|\right|#2\big)}
\newcommand{\expect}{\mathbb{E}}
\newcommand{\E}[1]{\expect\left[#1\right]}
\newcommand{\EE}[2]{\expect_{#1}\left[#2\right]}
\newcommand{\qz}[0]{\q{\z}}

\newcommand{\qzg}[1]{\q{\z|#1}}

\newcommand{\px}[0]{\p{\x}}
\newcommand{\pz}[0]{\p{\z}}

\newcommand{\psetx}[0]{\p{\setx}}

\newcommand{\pxgz}[0]{\p{\x|\z}} 
\newcommand{\psetxgz}[0]{\p{\setx|\z}}
\newcommand{\pzgsetx}[0]{\p{\z|\setx}}
\newcommand{\pzgx}[0]{\p{\z|\x}}

\newcommand{\LB}[0]{\mathcal{L}\left(\thetab, \phib, \x \right)}

\begin{document}

\maketitle

\begin{abstract}
The joint analysis of biomedical data in Alzheimer's Disease (AD) is important for better clinical diagnosis and to understand the relationship between biomarkers.
However, jointly accounting for heterogeneous measures poses important challenges related to the modeling of the variability and the interpretability of the results.
These issues are here addressed by proposing a novel multi-channel stochastic generative model.
We assume that a latent variable generates the data observed through different channels (\eg clinical scores, imaging, \ldots) and describe an efficient way to estimate jointly the distribution of both latent variable and data generative process.
Experiments on synthetic data show that the multi-channel formulation allows superior data reconstruction as opposed to the single channel one.
Moreover, the derived lower bound of the model evidence represents a promising model selection criterion.
Experiments on AD data show that the model parameters can be used for unsupervised patient stratification and for the joint interpretation of the heterogeneous observations.
Because of its general and flexible formulation, we believe that the proposed method can find important applications as a general data fusion technique. 
\end{abstract}

\section{Introduction}
Physicians investigate their patients' status through various sources of information that in this work we call \textit{channels}.
For Alzheimer's Disease (AD), for example, the anamnestic questionnaire, genetic tests and brain imaging modalities are channels providing specific, complementary, and sometimes overlapping views on the patient's state \cite{Dubois2014,Jack2018}.

Tackling a complex disease like AD requires to establish a link between heterogeneous data channels.
However, simple univariate correlation analyses are limited in modeling power, and are prone to false positives when the data dimension is high.
To overcome the limitations of mass-univariate analysis, more advanced methods, such as Partial Least Squares (PLS), Reduced Rank Regression (RRR), or Canonical correlation analysis (CCA) \cite{Hotelling1936} have successfully been applied in biomedical research \cite{Liu2014}, along with multi-channel \cite{Kettenring1971,Luo2015} and non-linear \cite{Huang2009,Andrew2013} variants.

A common drawback of standard multivariate methods is that they are not generative.
Indeed, their formulation consists in projecting the observations in a latent lower dimensional space in which they exhibit certain desired characteristics like maximum correlation (CCA), maximum covariance (PLS), minimum regression error (RRR);
however these methods are limited in providing information on how this latent representation is expressed in the observations \cite{Haufe2014a}.
Moreover, techniques for model comparison should be applied to select the best number of dimensions for the latent representation and avoid overfitting.
While cross-validation is the standard model validation procedure, this requires holding-out data from the original dataset, thus leading to data loss at the training stage.

We need generative models that can actually describe the direct influence of the latent space on the observations, and model selection techniques leveraging solely on training data.
\textit{Bayesian-CCA} \cite{Klami2013} actually goes in this direction:
it is a generative formulation of the CCA defined on a latent variable that captures the shared variation between data channels.
Moreover, the Bayesian formulation allows the use of probabilistic model comparison.
However, Bayesian-CCA may not scale well to large dimensions and several channels.

In this work we aim at addressing the current methodological limitations in multi-channel analysis.
By leveraging on the recent developments in efficient \textit{Variational Inference} in Bayesian modeling, we propose a novel multi-channel stochastic generative model for the joint analysis of multi-channel heterogeneous data.
Our hypothesis is that a latent variable $\z$ generates the heterogeneous data $\setx$ observed through different channels $C$.
In this work we propose an efficient way to estimate jointly the latent variable distribution and the data likelihood $\p{\setx|\z}$, and we also investigate a mean for Bayesian model selection.
Our work generalizes the \textit{Variational Autoencoder} \cite{Kingma2013} and the Bayesian-CCA, making possible to jointly model multiple channels simultaneously and efficiently.

The next sections of this paper are organized as follows.
In Section 2 we present the derivation of the multi-channel variational model and we describe a possible implementation with Gaussian distributions parametrized by linear functions.
In Section 3 we apply our method on a synthetic dataset, as well as on a real multi-channel Alzheimer's disease dataset, to test the descriptive and predictive properties of the model.
In the last section we provide our discussions and conclusions.

\section{Method}
\subsection{Multi-Channel Variational Inference}
Let $\x = \{\x_c\}_{c=1}^C$ be a single observation of a set of $C$ channels, where each $\x_c \in \realnumbers^{d_c}$ is a $d_c$-dimensional vector.
Also, let $\z \in \realnumbers^{l}$ denote the $l$-dimensional latent variable commonly shared by each $\x_c$.
We propose the following generative process:
\begin{equation}\label{eq:model}
\begin{aligned}
&\z \sim \pz \\  
&\x_c \sim \p{\x_c|\z,\thetab_c}  
\qquad \text{for} \quad c \quad \text{in} \quad 1 \ldots C
\end{aligned}
\end{equation}
where $\pz$ is a prior distribution for the latent variable and $\p{\x_c|\z,\thetab_c}$ is a likelihood distribution for the observations conditioned on the latent variable.
We assume that the likelihood functions belong to a distribution family $\mathcal{P}$ parametrized by $\thetab_c$.
When the distributions are Gaussians parametrized by linear transformations, the model is equivalent to the \textit{Bayesian-CCA} (\cf \cite{Klami2013}, Eq. 3).
In the scenario depicted so far, solving the inference problem allows the discovery of the common latent space from which the observed data in each channel is generated.
The solution to the inference problem is given by deriving the posterior $\p{\z|\setx,\thetaset}$, that is not always computable analytically.
In this case, \textit{Variational Inference} \cite{Blei2016} can be applied to compute an approximate posterior.
In our setting, variational inference is carried out by introducing probability density functions $\q{\z|\x_c,\phib_c}$ that are on average as close as possible to the true posterior in terms of Kullback-Leibler divergence:
\begin{equation}
\label{eq:objective}
\underset{q \in \mathcal{Q}}{arg\,min} \,\, \EE{c}{\KL{\q{\z|\x_c,\phib_c}}{\p{\z|\setx,\thetaset}}}
\end{equation}
where the approximate posteriors $\q{\z|\x_c,\phib_c}$ belong to a distribution family $\mathcal{Q}$ parametrized by $\phib_c$, and represent the view on the latent space that can be inferred from each channel $\xb_c$.
Practically, solving the objective in \eqnref{eq:objective} allows to use on average every $\q{\z|\x_c,\phib_c}$ to approximate the true posterior distribution.
It can be shown that the maximization of the model evidence $\p{\setx}$ is equivalent to the optimization of the evidence lower bound $\LB$:
\begin{equation}\label{eq:LB}
\begin{aligned}
&\LB = \frac{1}{C} \sum\limits_{c=1}^C \underbrace{\EE{\qzg{\x_c,\phib_c}}{\textstyle \sum_{i=1}^C \ln \p{\x_i|\z,\thetab_c}}}_{\text{cross-reconstruction of all $\x_i$ from $\x_c$}} - \KL{\qzg{\x_c,\phib_c}}{\pz} \\
&= \ln \underbrace{\p{\setx}}_{\text{Evidence}} - \underbrace{\EE{c}{\KL{\q{\z|\x_c,\phib_c}}{\p{\z|\setx,\thetaset}}} }_{\geq 0}\\
\end{aligned}
\end{equation}
It can be shown that maximizing $\LB$ is equivalent to solving the objective in \eqnref{eq:objective} (\cf \asuppmat).
Moreover, being the lower bound linked to the data evidence up to a positive constant, \eqnref{eq:LB} allows to test $\LB$ as a surrogate measure of $\p{\setx}$ for Bayesian model selection.
This formulation is valid for any distribution family $\mathcal{P}$ and $\mathcal{Q}$, and the complete derivation of \eqnref{eq:LB} is in the \asuppmat

\subsubsection{Comparison with variational autoencoder (VAE).}
Our model extends the VAE \cite{Kingma2013}: the novelty is in the cross-reconstruction term labeled in \eqnref{eq:LB}.
In case $C=1$ the model collapses to a VAE.
In the case $C>1$ the cross-term forces each channel to the joint decoding of the other channels.
For this reason, our model is different from a stack of independent VAEs.
The dependence between encoding and decoding  across channels stems from the joint approximation of the posteriors (Formula (\ref{eq:objective})).

\subsubsection{Optimization of the lower bound.}
The optimization starts with a random initialization of the generative parameters $\thetab$ and the variational parameters $\phib$.
The expectation in the first row of \eqnref{eq:LB} can be computed by sampling from the variational distributions $\qzg{\x_c,\phib_c}$ and, when the prior and the variational distributions are Gaussians, the Kullback-Leibler term can be computed analytically (\cf \cite{Kingma2013}, appendix 2.A).
The maximization of $\LB$ with respect to $\thetab$ and $\phib$ is efficiently carried out through minibatch stochastic gradient descent implemented with the backpropagation algorithm.
For each parameter, adaptive learning rates are computed with \textit{Adam} \cite{Kingma2014b}.

\subsection{Gaussian linear case}\label{sec:gausslin}
Model (\ref{eq:model}) is completely general and can account for complex non-linear relationships modeled, for example, through deep neural networks.
However, for simplicity of interpretation, and validation purposes, in the next experimental section we will restrict our multi-channel variational framework to the \textit{Gaussian Linear Model}.
This is a special case, analogous to Bayesian-CCA, where the members of the generative family $\mathcal{P}$ and variational family $\mathcal{Q}$ are Gaussians parametrized by linear transformations.
The parameters of these transformations are thus optimized by maximizing the lower bound.
We define the members of the generative family $\mathcal{P}$ as Gaussians whose first moments are linear transformations of the latent variable $\zb$, and the second moments are parametrized by a diagonal covariance matrix, such that
$\p{\x_c|\z,\thetab_c} = \Gaussof{\x_c|\mathbf{G}_c^{(\mu)} \zb, diag(\mathbf{g}_c^{(\sigma)})}$,
where $\mathbf{G}_c^{\left(\mu\right)} \in \mathbb{R}^{d_c \times l}$
and
$\mathbf{g}_c^{(\sigma)} \in \realnumbers^{d_c}$.
The elements of
$\thetab_c = \{\mathbf{G}_c^{(\mu)}, \mathbf{g}_c^{(\sigma)}\}$
are the generative parameters to be optimized for every channel.
We also define the members of variational family
$\mathcal{Q}$
to be Gaussians whose moments are linear transformation of the observations, such that
$\qzg{\x_c,\phib_c} = \Gaussof{\x_c|\mathbf{V}_c^{(\mu)} \xb_c, diag(\mathbf{V}_c^{(\sigma)} \xb_c)}$
where
$\mathbf{V}_c^{(\mu)} \in \realnumbers^{l \times d_c}$
and
$\mathbf{V}_c^{(\sigma)} \in \realnumbers^{l \times d_c}$.
The elements of
$\phib_c=\{\mathbf{V}_c^{(\mu)}, \mathbf{V}_c^{(\sigma)}\}$
are the variational parameters to be optimized for every channel.

\section{Experiments}
In this section we illustrate the performance of the method extensively tested on a large scale synthetic dataset,
and we provide a real case example by jointly analyzing multimodal brain imaging and clinical scores in AD data.

\subsection{Experiments on Linearly generated synthetic datasets}

\subsubsection{Data generation procedure.}
Datasets $\x = \left\{\x_c\right\}$ with $c = 1..C$ channels where created according to the following model:

\begin{equation}
\begin{aligned}
&\z \sim \GaussStdDim{l} \\
&\epsilonb \sim \GaussStdDim{d_c} \\
&\mathbf{G}_c = diag\left( \mathbf{R}_c \mathbf{R}_c^T \right)^{-1/2} \mathbf{R}_c \\
&\x_c = \mathbf{G}_c \z + \snr^{-1/2} \cdot \epsilonb
\end{aligned}
\label{eq:gen_model}
\end{equation}
where  for every channel $c$, $\mathbf{R}_c \in \realnumbers^{d_c \times l}$ is a random matrix with $l$ orthonormal columns (\ie\ $\mathbf{R}_c^T \mathbf{R}_c = \mathbf{I}_l$),
$\mathbf{G}_c$ is the linear generative law,
and $\snr$ is the signal-to-noise ratio.
It's easy to demonstrate that the diagonal elements of the covariance matrix of $\x_c$ are inversely proportional to $\snr$, \ie\,
$diag\left(\E{\x_c\x_c^T}\right) = (1 + \snr^{-1}) \mathbf{I}_{d_c}$.
Scenarios where generated by varying one-at-a-time the dataset attributes, as listed in \tabref{table:simul_params}.

\begin{table}[!htbp]
\caption{Dataset attributes, varied one-at-a-time in the prescribed ranges, and used to generate scenarios according to \eqnref{eq:gen_model}.}
\centering
\begin{tabular}{lll}
\toprule
Symbol & Attribute description          & Range / Iteration list\\
\midrule
$C$    & Total channels                 & 2, 3, 5, 10\\
$d_c$  & Channel dimension              & 4, 8, 16, 32, 500\\
$l$    & Latent space dimension         & 1, 2, 4, 10, 20\\
$S$    & Number of samples/observations & 50, 100, 1000, 10000\\
$\snr$ & Signal-to-noise ratio          & 100, 10, 1, 0.1\\
       & Replication number (re-initialize $\mathbf{R}_c$)  & 1, 2, 3, 4, 5\\  
\bottomrule
\end{tabular}
\label{table:simul_params}
\end{table}

\subsubsection{Results.}
\begin{figure}[!htbp]
\centering
\begin{subfigure}{\textwidth}
        \centering
        \includegraphics[width=\textwidth]{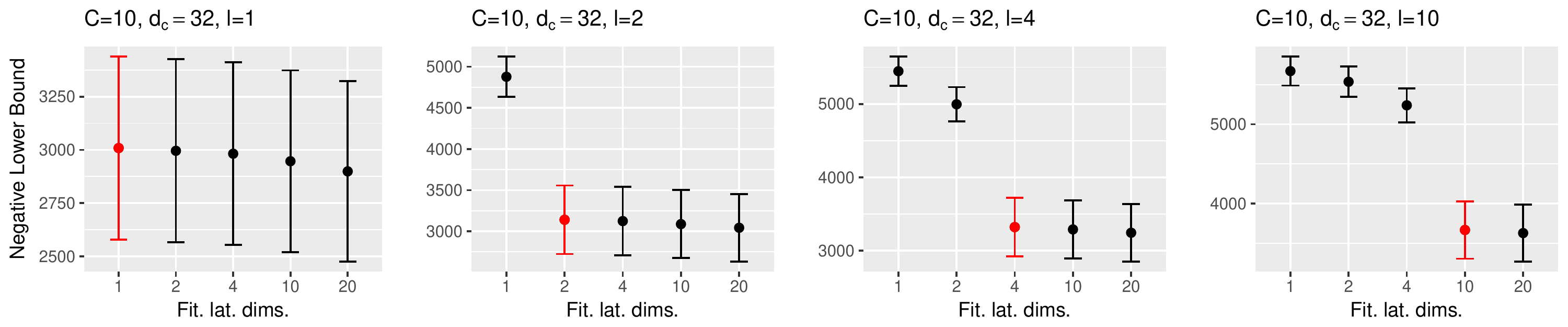}
	\caption{}
        \label{fig:model_sel_lin_ok}
\end{subfigure}
\begin{subfigure}{\textwidth}
        \centering
        \includegraphics[width=\textwidth]{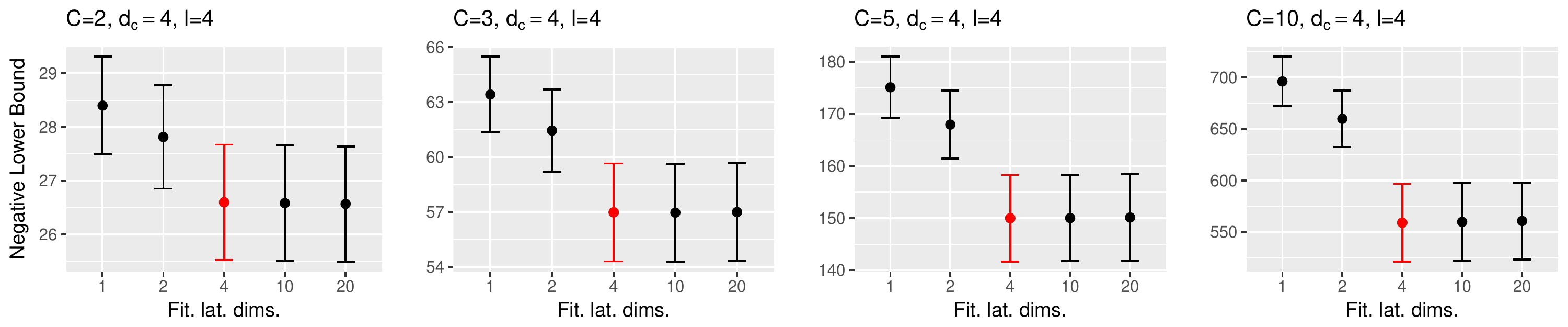}
	\caption{}
        \label{fig:model_sel_lin_C}
\end{subfigure}
\begin{subfigure}{\textwidth}
        \centering
        \includegraphics[width=\textwidth]{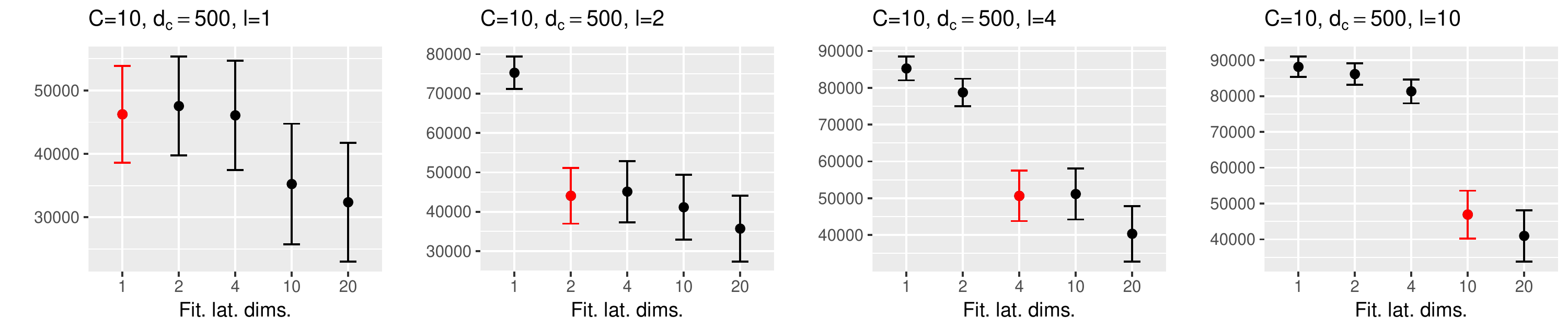}
	\caption{}
        \label{fig:model_sel_lin_overfit}
\end{subfigure}
\begin{subfigure}{\textwidth}
        \centering
        \includegraphics[width=\textwidth]{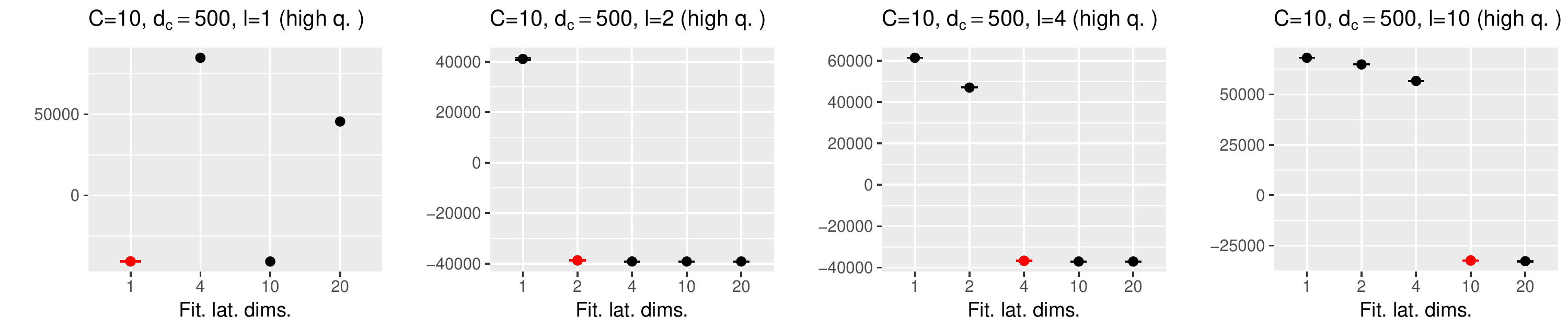}
	\caption{}
        \label{fig:model_sel_lin_hq}
\end{subfigure}
\caption{
Negative lower bound (NLB) on the synthetic training set computed at convergence for all the scenarios.
Each bar shows mean $\pm$ std.err. of $N = 80$ total experiments as a function of the number of fitted latent dimensions.
Red bars represents experiments where the number of true and fitted latent dimensions coincide.
(a) Experimental setup $C=10$, $d_c = 32$:
NLB  stops decreasing when the number of fitted latent dimension coincide with the generated ones;
notable gap between the under-fitted and over-fitted experiments (elbow effect).
(b) Experimental setup $d_c=4$ , $l=4$:
increasing the number of channels $C$ makes the elbow effect more pronounced.
(c) Experimental setup $C=10$ , $d_c=500$:
with high dimensional data ($d_c=500$) using the lower bound as a model selection criteria to assess the true number of latent dimensions may end up in overestimation.
(d) Restricted ($N = 5$ total experiments) high quality experimental setup $C=10$, $d_c = 500$, $S=10000$, $\snr=100$:
the risk to overestimate the true number of latent dimensions can be mitigated by increasing the $\snr$ and $S$ of the observations in the dataset.
}
\label{fig:model_sel_lin}
\end{figure}

At convergence, the loss function (negative lower bound) has a minimum when the number of fitted latent dimensions corresponds to the number of the latent dimensions used to generate the data, as depicted in \figref{fig:model_sel_lin_ok}.
When increasing the number of fitted latent dimensions, a sudden decrease of the loss (elbow effect) is indicative that the true number of latent dimensions has been found.
In \figref{fig:model_sel_lin_C} we show also that the elbow effect becomes more pronounced with increasing number of data channels.
Ambiguity in identifying the elbow, instead, may rise for high-dimensional data channels (\figref{fig:model_sel_lin_overfit}).
In these cases, increasing the sample size or the data quality in terms of $\snr$ can make the elbow point more noticeable (\figref{fig:model_sel_lin_hq}).

\begin{figure}[!htbp]
\centering
\begin{subfigure}{0.32\textwidth}
        \includegraphics[width=\textwidth]{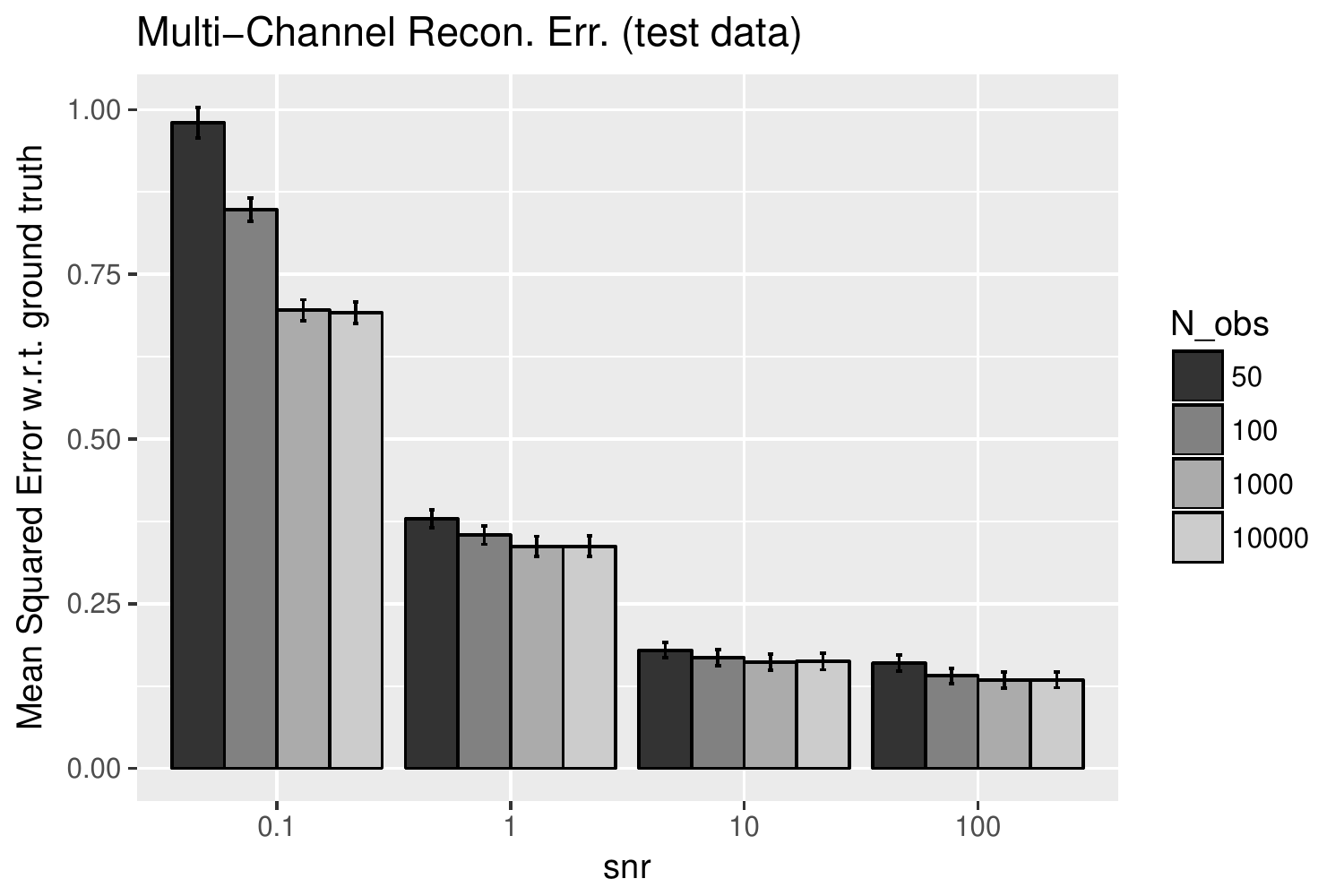}
        \caption{}
        \label{fig:recon_acc_multi_channel}
\end{subfigure}
\hfill
\begin{subfigure}{0.32\textwidth}
        \includegraphics[width=\textwidth]{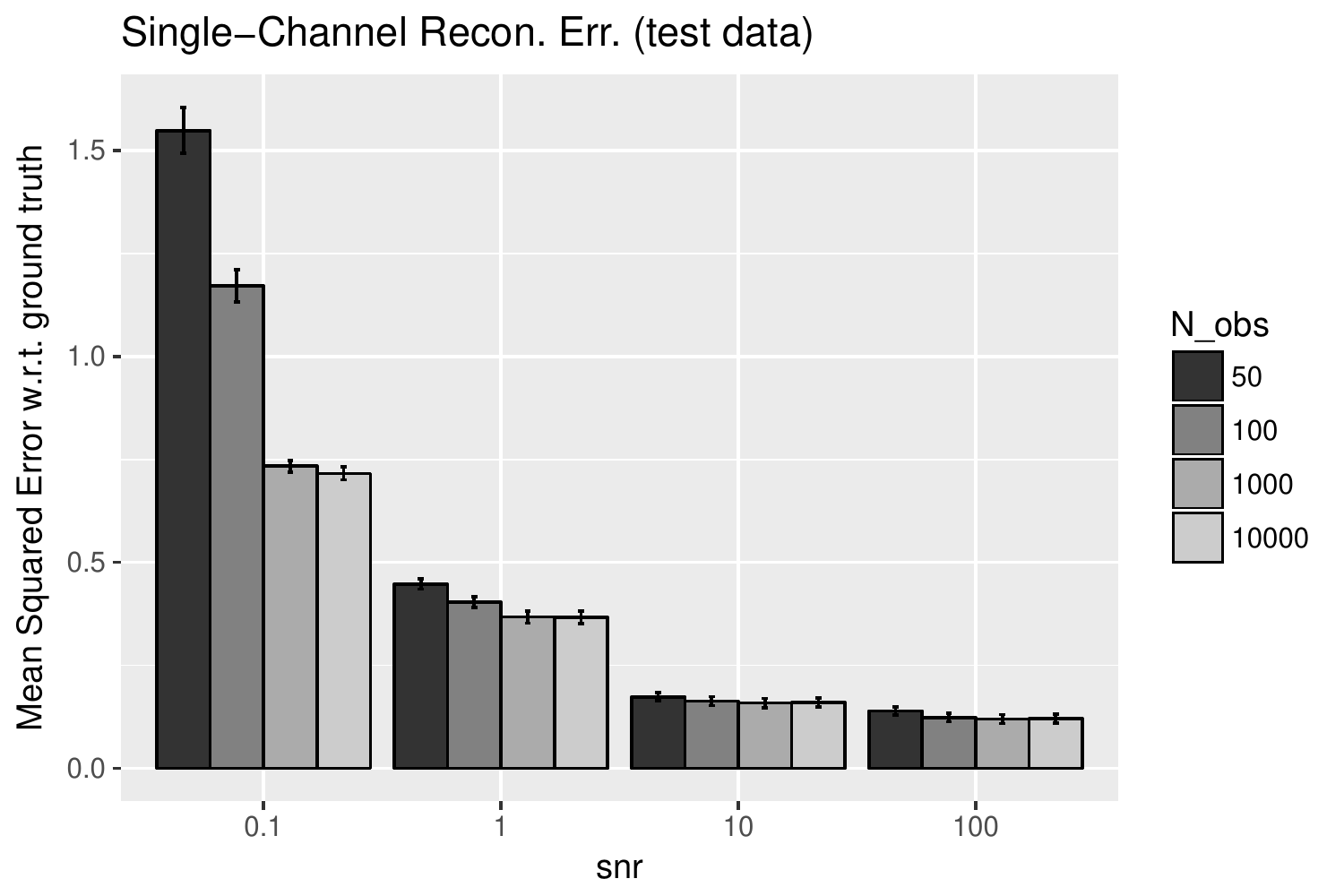}
        \caption{}
        \label{fig:recon_acc_single_channel}
\end{subfigure}
\hfill
\begin{subfigure}{0.32\textwidth}
        \includegraphics[width=\textwidth]{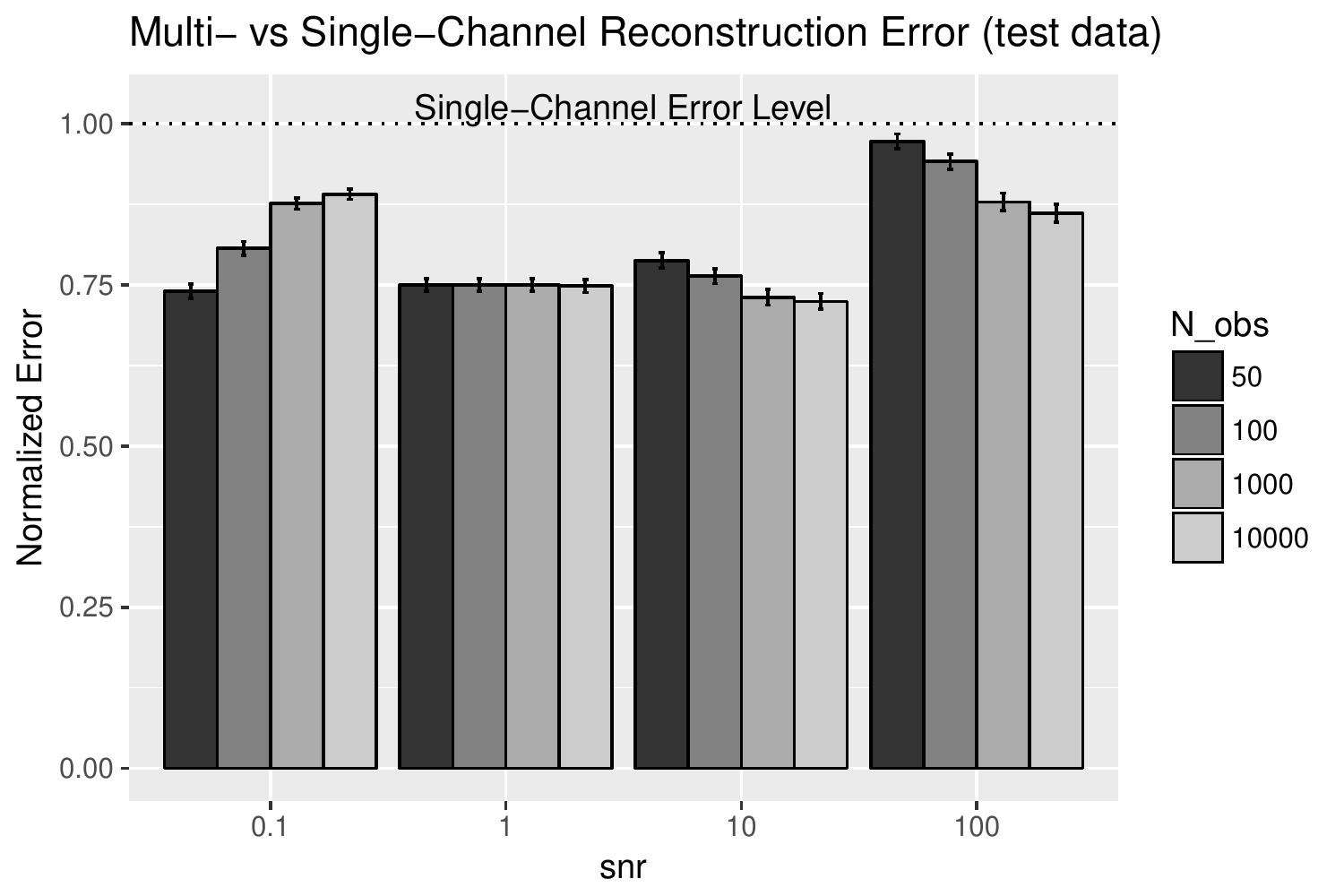}
        \caption{}
        \label{fig:recon_acc_ratio}
\end{subfigure}
\caption{
Reconstruction error on synthetic test data reconstructed with the multi-channel model.
The reconstruction is better for high $\snr$ and high training data sample size.
Scenarios where generated by varying one-at-a-time the dataset attributes listed in \tabref{table:simul_params} for a total of $8\,000$ experiments.
(a) Mean squared error from the ground truth test data using the Multi-Channel reconstruction: $\hat{\x}_i = \EE{c}{\EE{\qzg{\x_c}}{\p{\x_i|\z}}}$
(b) Mean squared error from the ground truth test data using the Single-Channel reconstruction: $\hat{\x}_i = \EE{\qzg{\x_i}}{\p{\x_i|\z}}$
(c) Ratio between Multi- \textit{vs} Single-Channel reconstruction errors: we notice that the error made in ground truth data recovery with multi-channel information is systematically lower than the one obtained with a single-channel decoder.
}
\label{fig:recon_acc}
\end{figure}

Concerning the reconstruction performance on test data, we observed that the performance of the model increases with higher \snr, sample size, and number of channels (\figref{fig:recon_acc}).
The reconstruction of a channel $i$ can be done by applying the decoder $i$ to the latent variables generated from all the channels and then average the results, according to the Formula:

\begin{equation}
\hat{\x}_i = \EE{c}{\EE{\qzg{\x_c}}{\p{\x_i|\z}}}
\label{eq:recon}
\end{equation}
The number of channels $N_c$ used in the reconstruction of channel $i$ can vary from $1$ to $C$.
In case $N_c=1$, the decoder of the channel $i$ is applied to the latent variable inferred from the same channel, similarly to a single channel VAE, with the difference that the cost function used to train the model, provided in the first row of \eqnref{eq:LB}, still takes in account all the channels.
We notice that the error made in ground truth data recovery with multi-channel information (case $N_c=C$) is systematically lower than the one obtained with a single-channel decoder (\figref{fig:recon_acc}).

\subsection{Application to clinical and medical imaging data in AD}

\subsubsection{Data preparation.}
Data used in the preparation of this article were obtained from the Alzheimer's Disease Neuroimaging Initiative (ADNI) database (\href{http://adni.loni.usc.edu}{adni.loni.usc.edu}).
The ADNI was launched in 2003 as a public-private partnership, led by Principal Investigator Michael W. Weiner, MD.
For up-to-date information, see \href{www.adni-info.org}{www.adni-info.org}.

We fit our model with linear parameters to clinical imaging channels acquired on $504$ subjects.
The clinical channel is composed of six continuous variables generally recorded in memory clinics (age, mini-mental state examination, adas-cog, cdr, faq, scholarity);
the three imaging channels are structural MRI (gray matter only), functional FDG-PET, and Amyloid-PET, each of them composed by continuous measures averaged over $90$ brain regions mapped in the AAL atlas \cite{Tzourio-Mazoyer2002a}.
Raw data from the imaging channels where coregistered in a common geometric space, and visual quality check was performed to exclude registration errors.
Data was centered and standardized across dimensions.
Model selection was carried out by comparing the lower bound for several fitted latent dimensions.

\subsubsection{Results.}
\begin{figure}[!htbp]
\centering
\begin{minipage}{0.54\textwidth}
\begin{subfigure}{\textwidth}
	\includegraphics[width=0.90\textwidth]{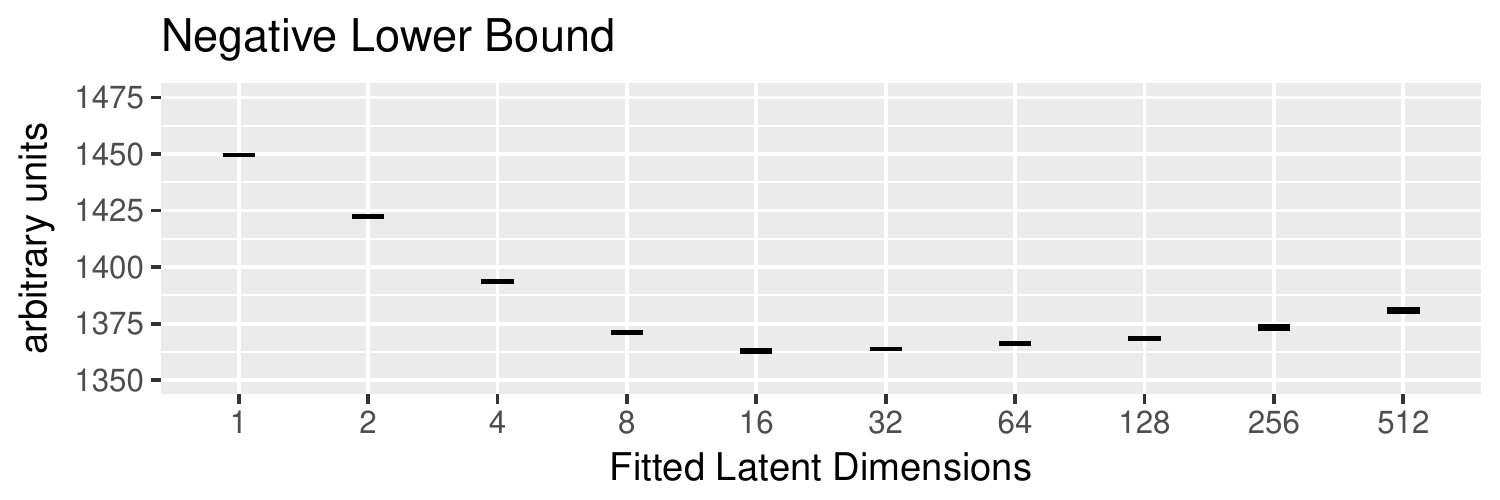}
	\caption{}
	\label{fig:adni_model_sel_a}
\end{subfigure}
\vfill
\begin{subfigure}{\textwidth}
	\includegraphics[width=\textwidth]{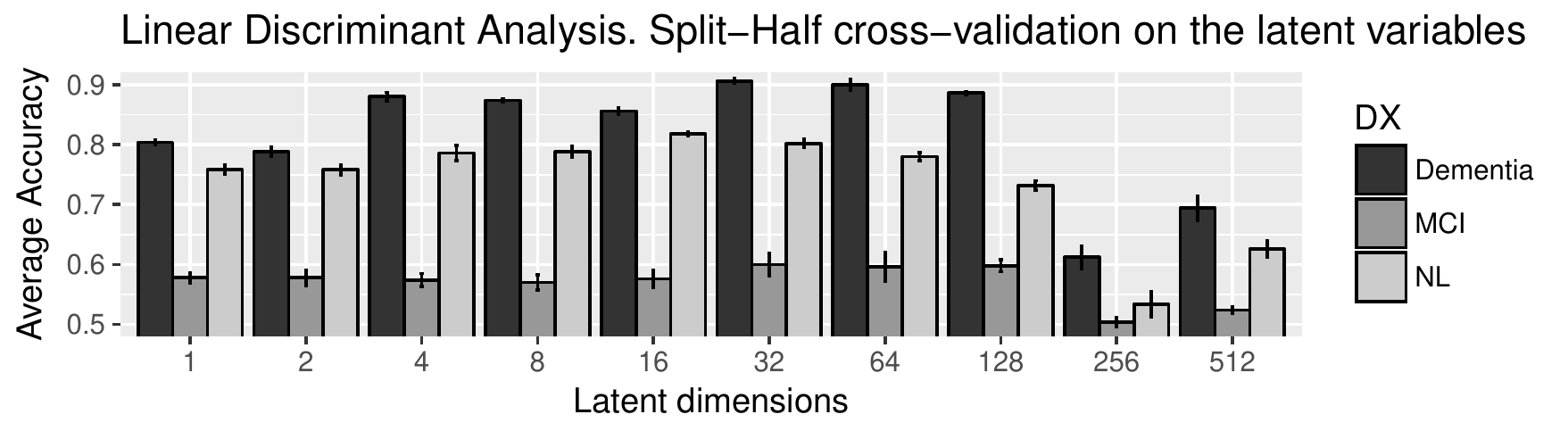}
	\caption{}
	\label{fig:lda_adni}
\end{subfigure}
\end{minipage}
\hfill
\begin{subfigure}{0.45\textwidth}
	\centering
	\includegraphics[width=\textwidth]{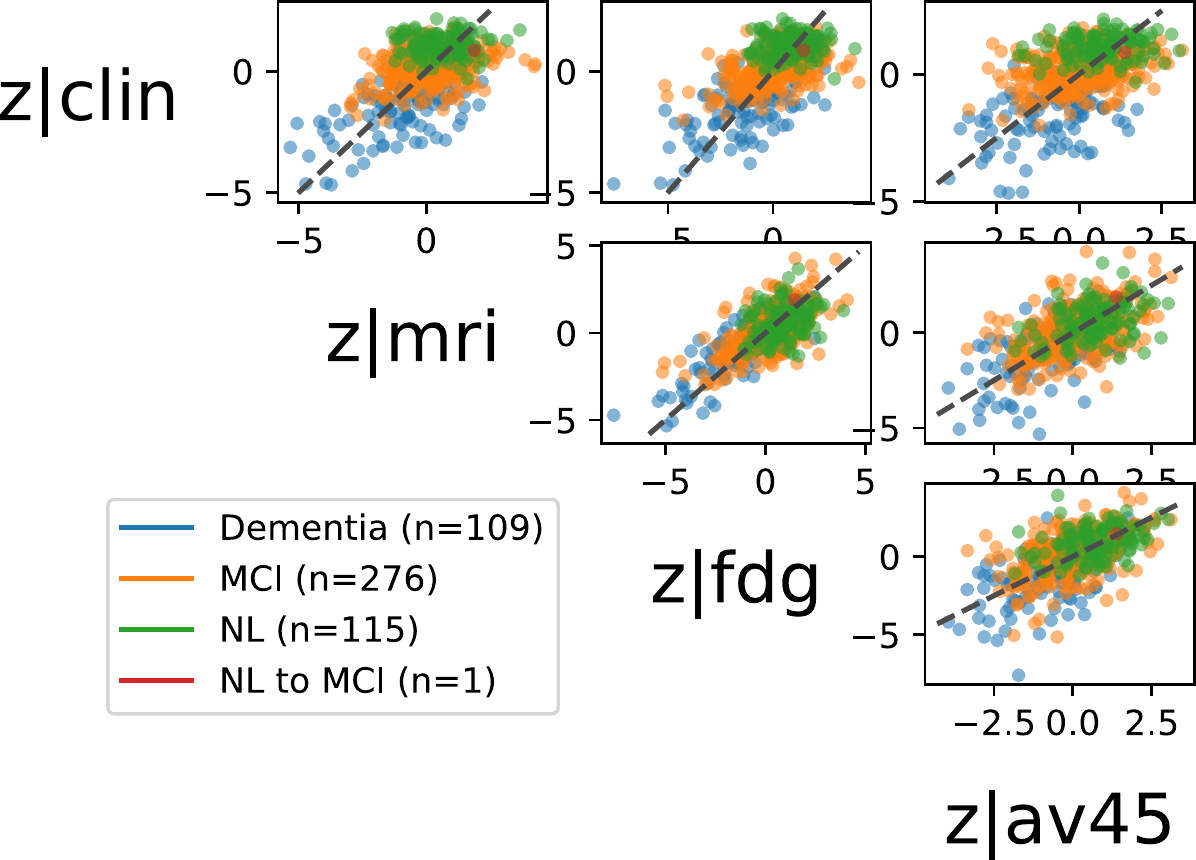}
	\caption{}
	\label{fig:adni_latent_dim}
\end{subfigure}
\caption{
Modeling results on ADNI data.
(a) The negative lower bound has a minimum when fitting 16 latent dimensions.
(b) Classification performance of the models: maximum accuracy for classes identification occurs with 16 and 32 lat. dims., in agreement with (a).
(c) Pairwise representations of one latent dimension (out of $16$) inferred from each of the four data channel.
Although the optimization is not supervised to enforce clustering, subjects appear stratified by disease classes.
} 
\label{fig:adni_model_sel}
\end{figure}

As depicted in \figref{fig:adni_model_sel_a}, we found that model selection through the lower bound identifies in a range around $16$ the number of latent dimensions that optimally describe the observations.
When fixing $16$ latent dimensions, in the latent space subjects appear stratified by disease status, an information that was not directly introduced ahead.
This is shown for one latent dimension in \figref{fig:adni_latent_dim}.
For each model fitted with increasing latent dimensions, the classification accuracy in predicting the disease status was assessed through split-half cross-validation linear discriminant analysis on the latent variables (\figref{fig:lda_adni}).
Maximum accuracy for disease classification occurs at $16$ and $32$ latent dimensions, an optimum location also identified through the lower bound.

\begin{figure}[!htbp]
\centering
\begin{subfigure}{0.45\textwidth}
        \includegraphics[width=\textwidth]{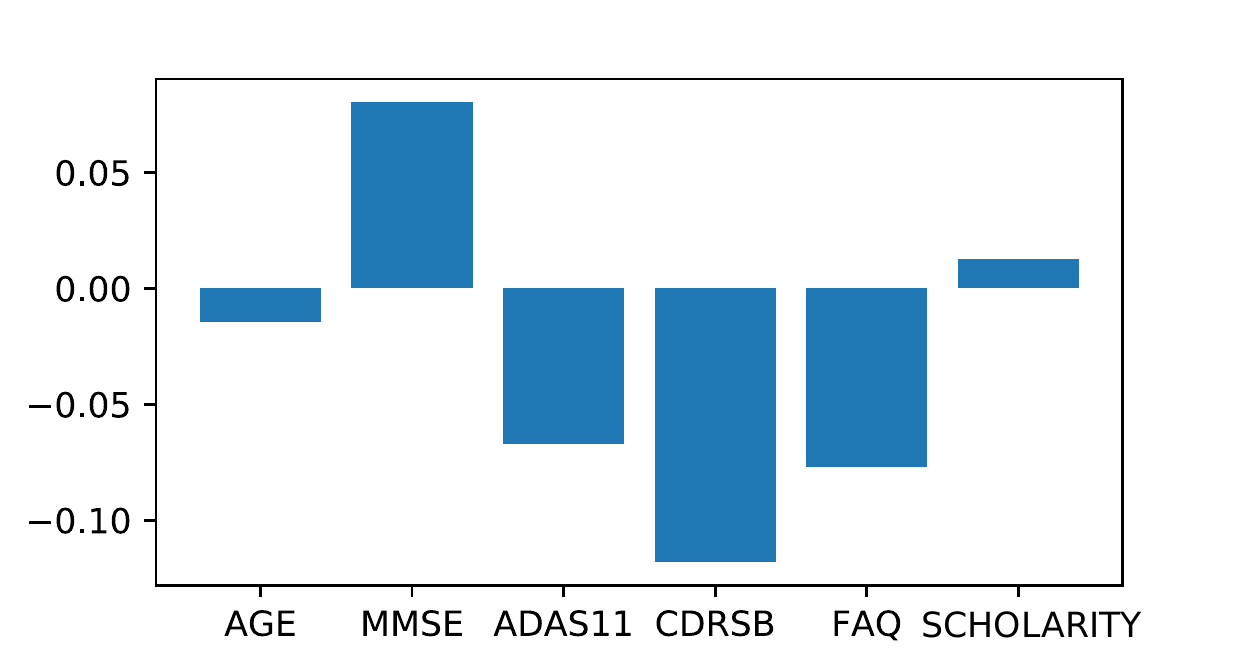}
\end{subfigure}
\begin{subfigure}{0.15\textwidth}
        \includegraphics[width=\textwidth]{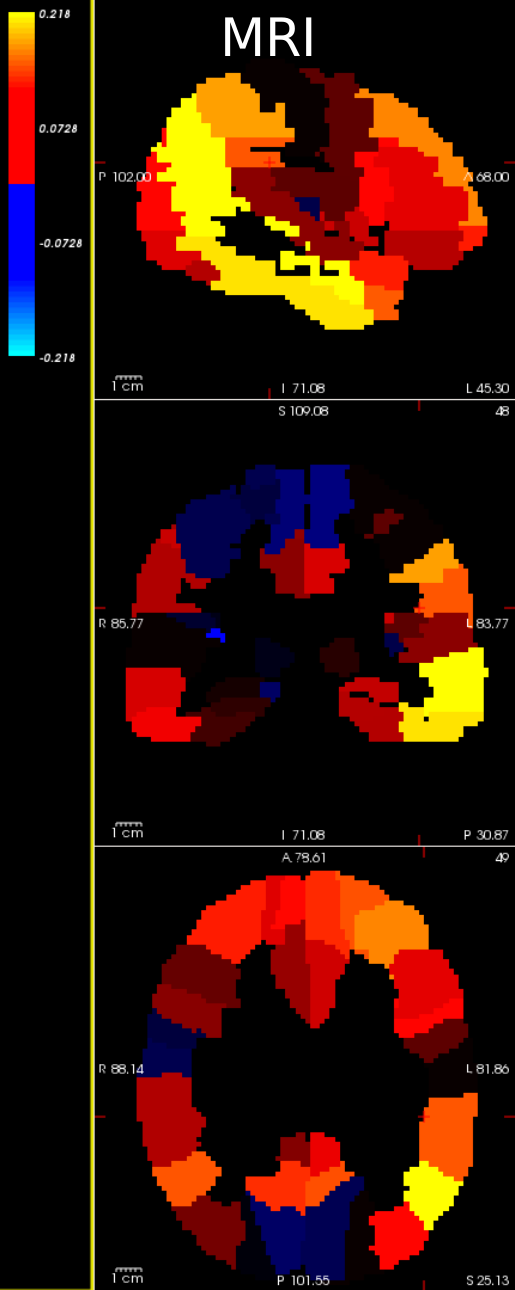}
\end{subfigure}
\begin{subfigure}{0.15\textwidth}
        \includegraphics[width=\textwidth]{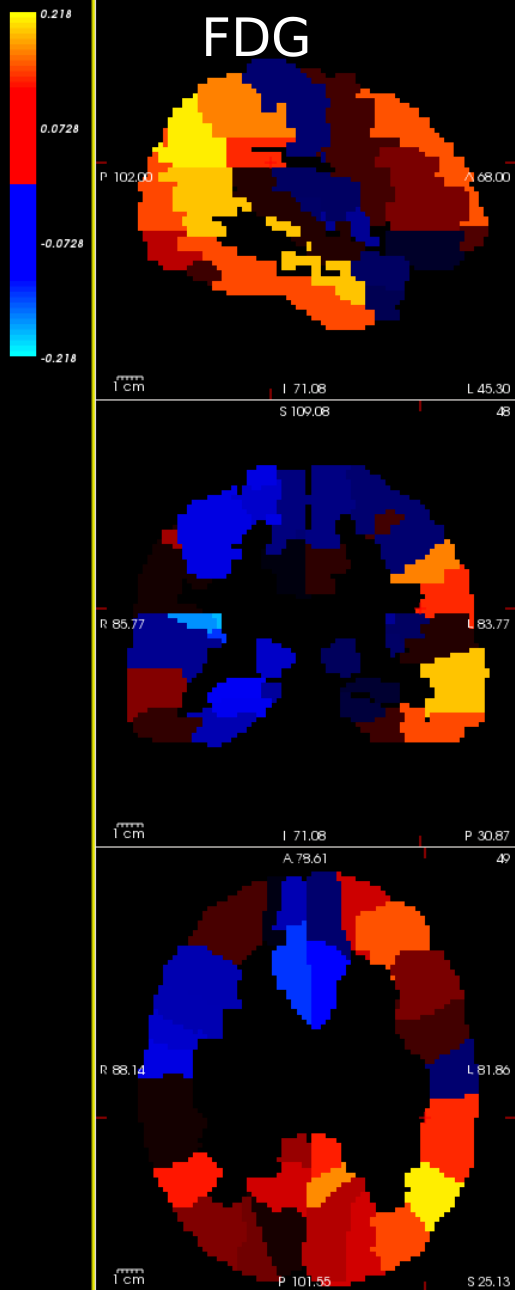}
\end{subfigure}
\begin{subfigure}{0.15\textwidth}
        \includegraphics[width=\textwidth]{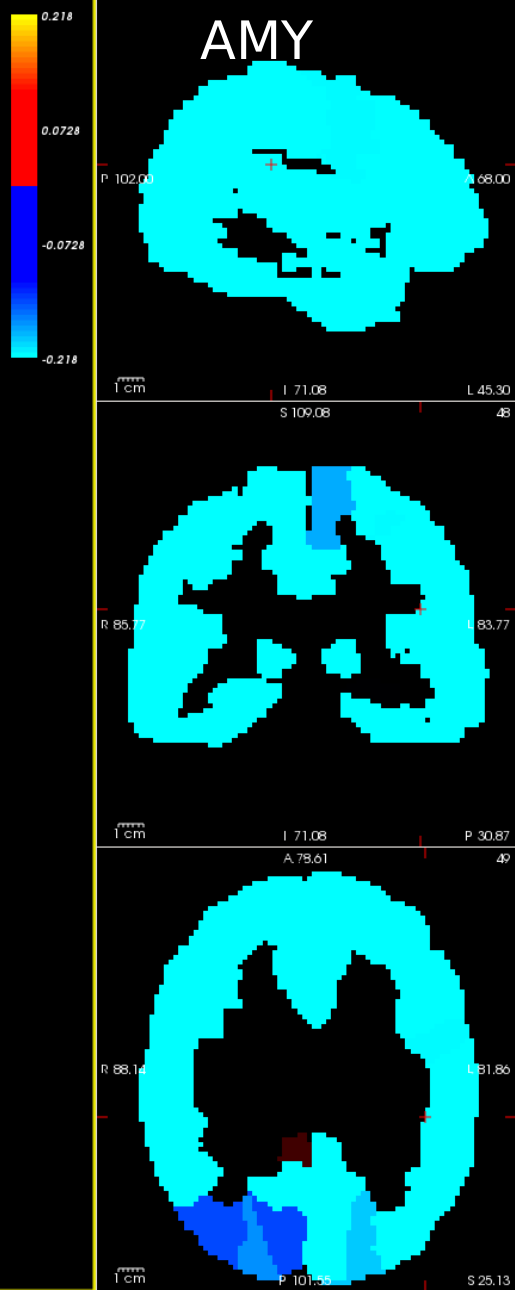}
\end{subfigure}
\caption{
Generative parameters $\phib_c^{(\mu)}$ of the four channels associated to the latent dimension in \figref{fig:adni_latent_dim}. 
The clinical parameters are age, mini-mental state examination (mmse), adas-cog (adas11), cdr-sb, faq, scholarity.
In the imaging channels, red is used for positive parameters, blue for negative ones.
} 
\label{fig:adni_gen_pars}
\end{figure}

\figref{fig:adni_gen_pars} shows the generative parameters $\phib_c$ of the four channels associated to the latent dimension shown in \figref{fig:adni_latent_dim}.
The generative parameters describe a plausible relationship between this latent dimension and the heterogeneous observations in the ADNI dataset, coherently with the research literature on Alzheimer's Disease, e.g. low amyloid deposition, high mini-mental state examination score, low adas-cog score, low cdr \cite{Murphy2010,Doraiswamy2012}, etc.

\section{Discussion and Conclusion}
We presented a multi-channel stochastic framework based on a probabilistic generative formulation. 
The performance of our multi-channel model was shown in the case of Gaussian distributions with moments parametrized by linear functions.
In the real case scenario of AD modeling, the model allowed the unsupervised stratification of the latent variable by disease status, providing evidence for a physiological interpretation of the latent space.
The generative parameters can therefore encode clinically meaningful relationships across multi-channel observations.
Although the use of the lower bound for model selection presents theoretical limitations \cite{Blei2016}, we found that it leads to good approximation of the marginal likelihood, thus providing a basis for model selection.

Future extension of this work will concern model with non-linear parameterization of the distributions, easily implementable through deep neural networks.
The use of non-Gaussian distributions can also be tested.
Given the scalability of our variational model, application to high resolution images may be also easily implemented.
To increase the model classification performance, supervised clustering of the latent space will be introduced, for example, by adding an appropriate cost function to the lower bound.
Also, introducing sparsity to remove redundancies may ease the identification and interpretation of the most informative parameters.
Lastly, due to the general formulation, the proposed method can find various applications as a general data fusion technique, not limited to the biomedical research area.

\subsection*{Acknowledgments}
\label{sec:ack}

This work has been supported by:
\begin{itemize}
\item the French government, through the UCA\textsuperscript{JEDI} Investments in the Future project managed by the National Research Agency (ANR) with the reference number ANR-15-IDEX-01;
\item the grant AAP Santé 06 2017-260 DGA-DSH, and by the Inria Sophia Antipolis - Méditerranée, "NEF" computation cluster.
\item Data collection and sharing for this project was funded by the Alzheimer's Disease Neuroimaging Initiative (ADNI) (National Institutes of Health Grant U01 AG024904) and DOD ADNI (Department of Defense award number W81XWH-12-2-0012).
ADNI is funded by the National Institute on Aging, the National Institute of Biomedical Imaging and Bioengineering, and through generous contributions from the following: AbbVie, Alzheimer's Association; Alzheimer's Drug Discovery Foundation; Araclon Biotech; BioClinica, Inc.; Biogen; Bristol-Myers Squibb Company; CereSpir, Inc.; Cogstate; Eisai Inc.; Elan Pharmaceuticals, Inc.; Eli Lilly and Company; EuroImmun; F. Hoffmann-La Roche Ltd and its affiliated company Genentech, Inc.; Fujirebio; GE Healthcare; IXICO Ltd.; Janssen Alzheimer Immunotherapy Research \& Development, LLC.; Johnson \& Johnson Pharmaceutical Research \& Development LLC.; Lumosity; Lundbeck; Merck \& Co., Inc.; Meso Scale Diagnostics, LLC.; NeuroRx Research; Neurotrack Technologies; Novartis Pharmaceuticals Corporation; Pfizer Inc.; Piramal Imaging; Servier; Takeda Pharmaceutical Company; and Transition Therapeutics.
The Canadian Institutes of Health Research is providing funds to support ADNI clinical sites in Canada.
Private sector contributions are facilitated by the Foundation for the National Institutes of Health (\url{www.fnih.org}).
The grantee organization is the Northern California Institute for Research and Education, and the study is coordinated by the Alzheimer's Therapeutic Research Institute at the University of Southern California.
ADNI data are disseminated by the Laboratory for NeuroImaging at the University of Southern California.
\end{itemize}

\bibliographystyle{splncs04}
\bibliography{tex/paper_biblio}

\newpage
\section*{Supplementary Material}
\label{sec:supmat}

\subsection*{Derivation of the Lower Bound}
In the following derivation we denote $\x = \{\setx\}$ to leave the notation uncluttered.
For the same reason we will omit the variational and generative parameters $\phib$ and $\thetab$.

Variational inference is carried out by introducing a set of probability density functions $\q{\z|\x_c}$, belonging to a distribution family $\mathcal{Q}$, that are on average as close as possible to the true posterior over the latent variable $\p{\z|\x}$.
In other words we aim to solve the following minimization problem:

\begin{equation}\label{eq:min_true_posterior}
\underset{q \in \mathcal{Q}}{arg\,min} \,\, \EE{c}{\KL{\q{\z|\x_c}}{\pzgsetx}}
\end{equation}
Given the intractability of $\p{\z|\x}$ for most complex models, we cannot solve directly this optimization problem.
We look then for an equivalent problem, by rearranging the objective:

\begin{equation}
\begin{aligned}
&\EE{c}{\KL{\qzg{\x_c}}{\p{\z|\x}}} = \EE{c}{\int_{\z} \qzg{\x_c} \big( \ln \qzg{\x_c} - \ln \pzgx \big)\,\dz} \\
                                   &= \EE{c}{\int_{\z} \qzg{\x_c} \big( \ln \qzg{x_c} - \ln \pxgz - \ln \pz + \ln \px \big)\,\dz} \\
                                   &= \ln \px + \EE{c}{\KL{\qzg{\x_c}}{\pz} - \EE{\qzg{\x_c}}{\ln \pxgz}} \\
\end{aligned}
\end{equation}
where in the middle line we use the Bayes' theorem to factorize the true posterior $\pzgx$.
Now, we can reorganize the terms, such that:

\begin{equation}\label{eq:exact_evidence}
\begin{aligned}
&\ln \px - \underbrace{ \EE{c}{\KL{\qzg{\x_c}}{\p{\z|\x}}} }_{\geq 0} = \\
&= \underbrace{ \EE{c}{ \EE{\qz}{\ln \psetxgz} - \KL{\qzg{\x_c}}{\pz} } }_{\text{lower bound }\mathcal{L}}
\end{aligned}
\end{equation}
Since the KL term in the \lhs\ is always non-negative, the \rhs\ is a lower bound to the log evidence.
Thus, by maximizing the lower bound we also maximize the data log evidence while solving the minimization problem in (\ref{eq:min_true_posterior}).

The hypothesis that every channel is conditionally independent from all the others given $\z$, allows to factorize the data likelihood as $\p{\setx|\zb} = \prod_{i=1}^C \p{\x_i|\z}$, so that the lower bound becomes:

\begin{equation}
\mathcal{L} =  \EE{c}{ \EE{\qzg{\x_c}}{\textstyle \sum_{i=1}^C \ln \p{\x_i|\z}} - \KL{\qzg{\x_c}}{\pz} }
\end{equation}
Finally, assuming every channel is equally likely to be observed with probability $1/C$, we can rewrite equation~(\ref{eq:exact_evidence}) as:
\begin{equation}
\ln \psetx \geq \frac{1}{C} \sum\limits_{c=1}^C \EE{\qzg{\x_c}}{\textstyle \sum_{i=1}^C \ln \p{\x_i|\z}} - \KL{\qzg{\x_c}}{\pz}
\end{equation}


\begin{figure}[!htbp]
\centering
\begin{subfigure}{\textwidth}
        \centering
        \includegraphics[width=\textwidth]{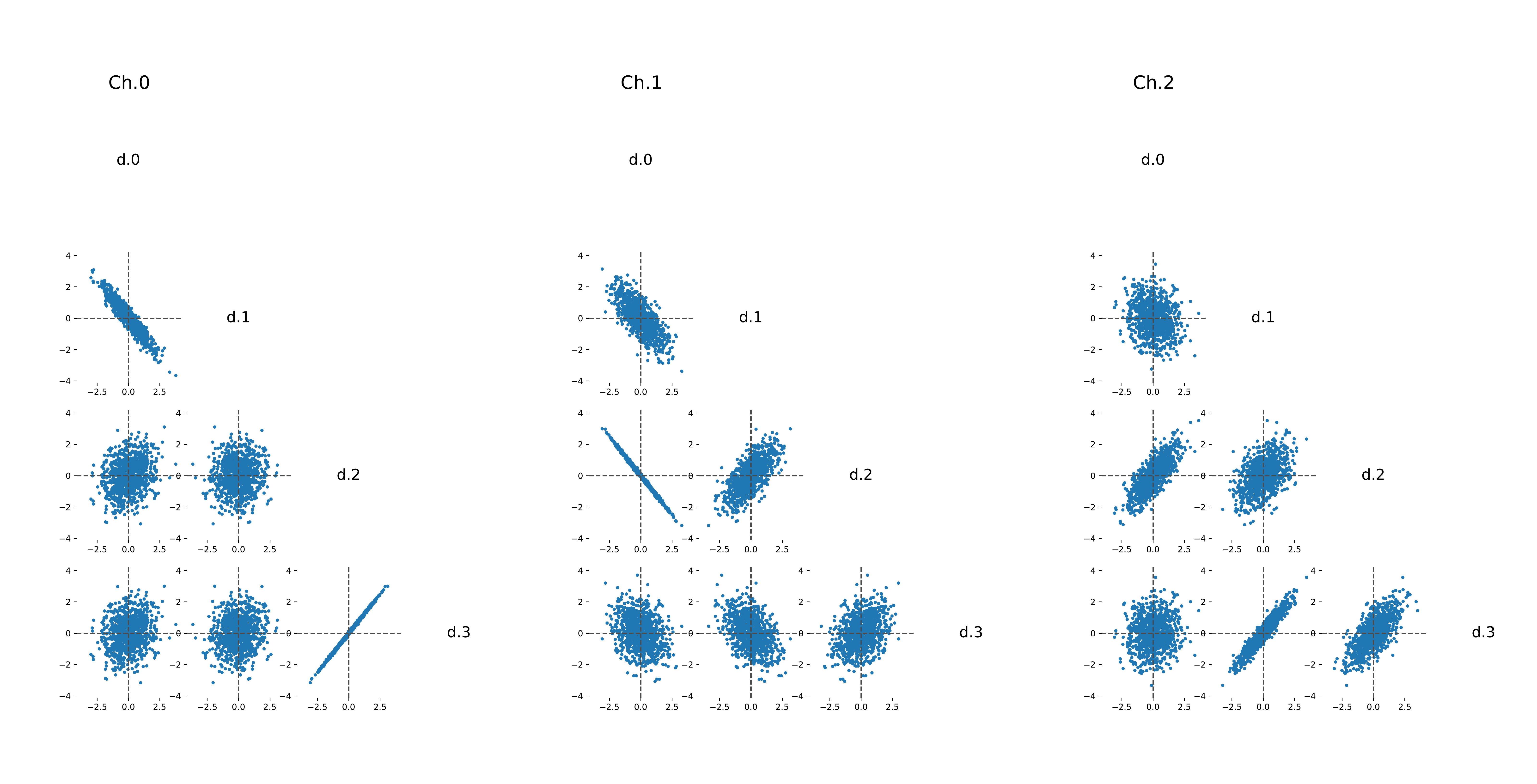}
	\caption{Ground truth}
        \label{fig:gen_lin_gt}
\end{subfigure}
\begin{subfigure}{\textwidth}
        \centering
        \includegraphics[width=\textwidth]{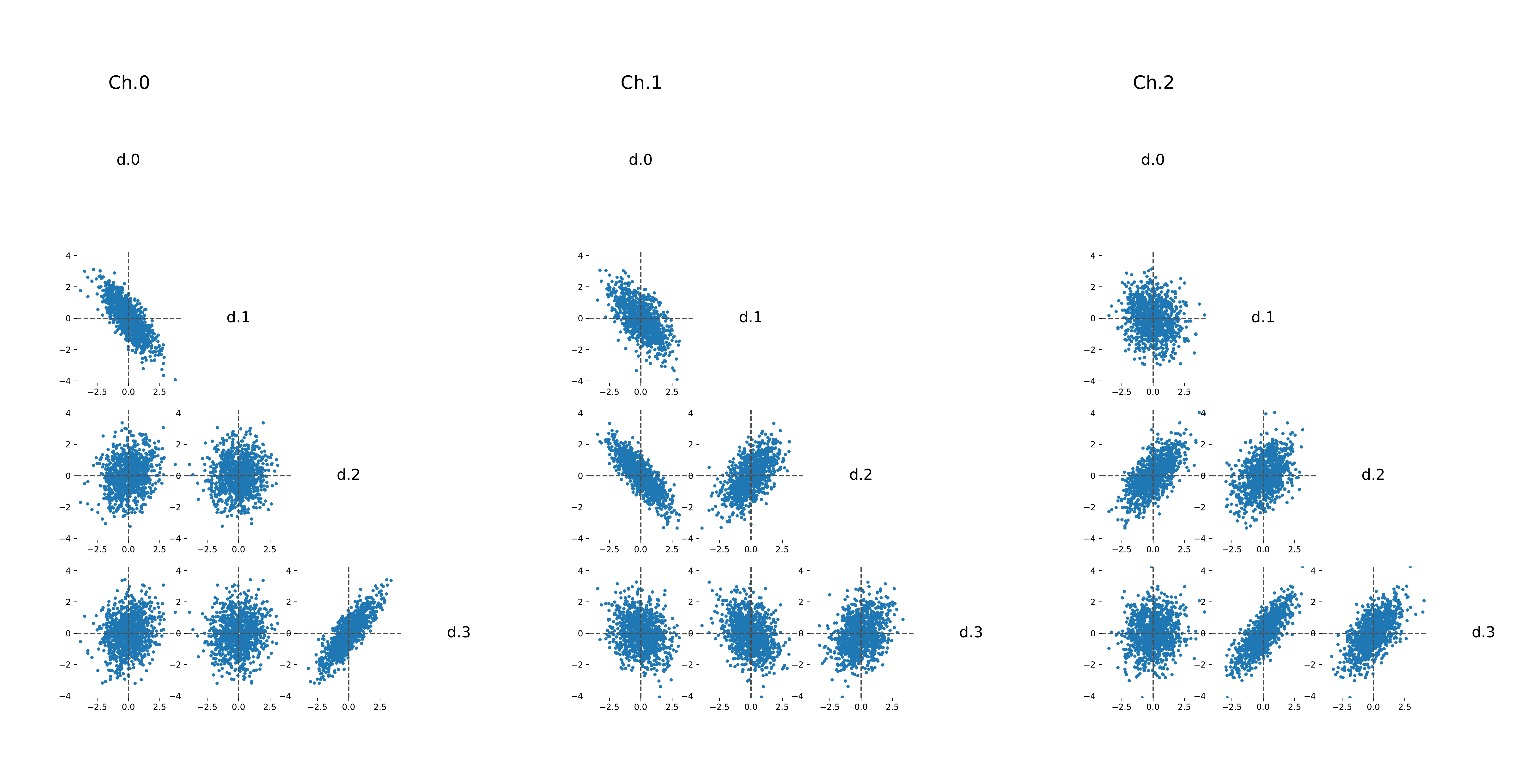}
	\caption{Noisy observations ($\snr=5$)}
        \label{fig:gen_lin_obs}
\end{subfigure}
\begin{subfigure}{\textwidth}
        \centering
        \includegraphics[width=\textwidth]{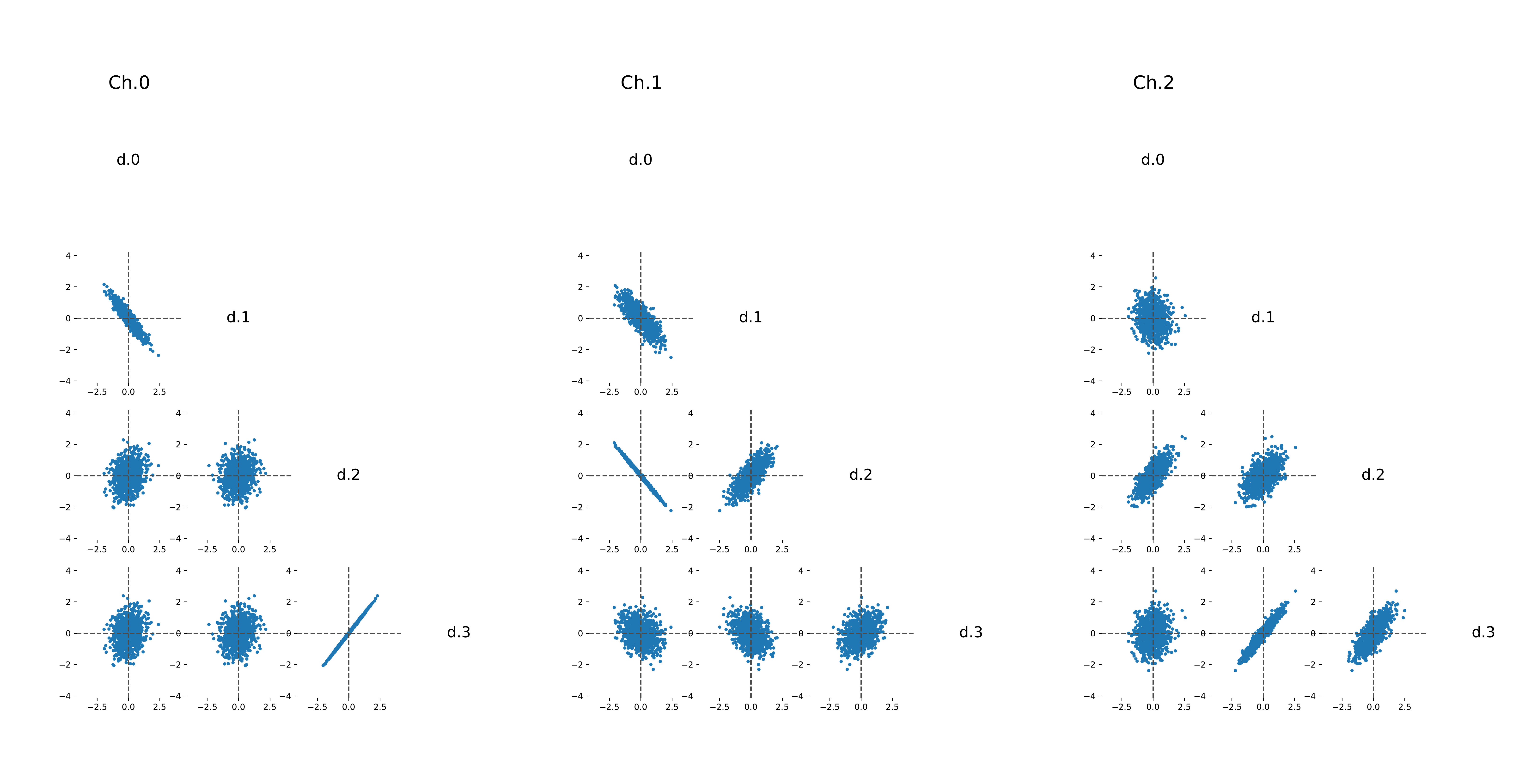}
	\caption{Reconstruction}
        \label{fig:gen_lin_recon}
\end{subfigure}
\caption{
Pairwise representation of the four dimensions $d$ of three data channels $Ch$, generated from a two-dimensional latent dimension $\zb \sim \Gausstd$, according to \eqnref{eq:gen_model}.
Noisy data was fitted with our model with a linear reparameterization.
(a) Ground truth ($\snr=0$).
(b) Observations used to fit the multi-channel model ($\snr=5$).
(c) Data generated from the latent variable inferred from the noisy data.
}
\label{fig:gen_lin}
\end{figure}

\end{document}